# Bouncing of a droplet on a superhydrophobic surface in AC electrowetting

Seung Jun Lee, Sanghyun Lee, and Kwan Hyoung Kang

Department of Mechanical Engineering, Pohang University of Science and

Technology, San 31, Pohang 790-784, South Korea

October 16, 2009

**Abstract**

We introduce a droplet-jumping phenomenon on a superhydrophobic surface driven by the resonant AC electrowetting. The resonant electrical actuation enables a droplet to accumulate sufficient surface energy for jumping, and superhydrophobic surface minimizes adhesion and hysteresis effects. They provide the effective energy conversion from the surface energy to the kinetic energy and improve the stability and the reproducibility of the droplet jumping. The controlled droplet jumping made by the resonant AC electrowetting could place another milestone in digital microfluidics by establishing a potential way to realize the three-dimensional droplet manipulation based on the conventional single plate EWOD configurations. This abstract is related to a fluid dynamics video for the gallery of fluid motion 2009.

Digital microfluidics based on the droplet manipulation by the electrowetting [1] enables microfluidic biochemical processes to be performed in an isolated droplet. The scalable and reconfigurable droplet manipulation of digital microfluidics, however, is primarily limited in two-dimensions. The increasing demand for the three-dimensional (3D) configuration in microfluidics can be also applicable to the digital microfluidics to broaden the scope of applications through 3D manipulation of droplets. In addition, the 3D droplet manipulation could also provide solutions to the difficult problems in current planar digital microfluidics such as the cross-contamination by solute adsorption and the degradation of electrodes [2]. Although there are a few interesting approaches enabling droplet elevation such as the droplet translation on curved electrodes [3], which is rigorously two-dimensional, complete 3D droplet manipulation based on EWOD manipulation is still unprecedented. Controlled droplet jumping we introduce here is a potential approach for the fully 3D droplet manipulation, allowing droplets to separate from the electrode surface and to be transported to the target place. The movie shows controlled jumping of a sessile droplet from a superhydrophobic surface driven by the resonant AC electrowetting. The resonant electrical actuation is, thus, the key for the sessile droplet to accumulate sufficient energy for jumping; when on non-superhydrophobic surfaces, droplets could split due to the considerable adhesion competing against the accumulated kinetic energy by the resonant AC electrowetting for jumping [4]. After jumping, the droplet bounces on the superhydrophobic surface, where the minimized adhesion and hysteresis make the decay of the bouncing height considerably slow. Besides, superhydrophobic surfaces provide the effective energy conversion from the surface energy to the kinetic energy and improve the stability and the reproducibility of the droplet jumping. The stability and the reproducibility of superhydrophobic surfaces used in this demonstration [5] are found fairly acceptable for the controlled jumping of a droplet on demand (CJDD). The controlled droplet jumping made by the resonant AC electrowetting could place another milestone in digital microfluidics by establishing a potential way to realize the 3D digital microfluidics based on the current single plate EWOD configurations.

This work was supported by the Korea Research Foundation Grant funded by the Korean Government (MOEHRD, Basic Research Promotion Fund) (KRF-2006-331-D00058). The superhydrophobic substrates were supplied by Applasma Co., Ltd.

Videos can be found with the following links:
- [Video 1](#) – Low res. and
- [Video 2](#) – High res..

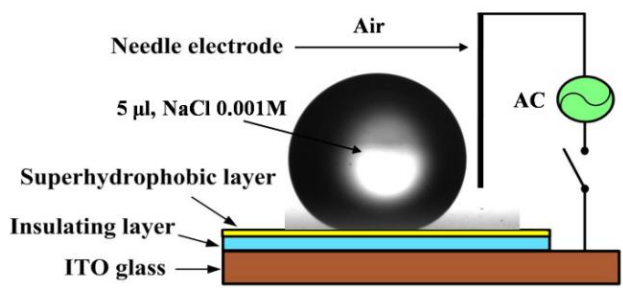

**Figure 1.** Schematic for the experimental configuration. The superhydrophobic base substrate is produced by the atmospheric rf plasma treatment (SHP 1000, App. Co., Ltd., www.applasma.com), which is finally performed on the parylene-C insulation layer (5 μm in thickness). The stainless steel needle electrode of 80 μm in diameter is the upper electrode and the ITO-coated glass is the bottom electrode. The ionic strength and the volume of the droplet shown in the movie are $10^{-3}$ M (NaCl) and 5 μL, respectively. To best match the resonance of the droplet, the frequencies of the driving potentials should be slightly different; 39Hz is for $100V_{RMS}$ and 42Hz for $90V_{RMS}$, because bigger amplitude requires longer period. The jumping and bouncing motions of a droplet are captured by a high-speed camera at 3000 fps.

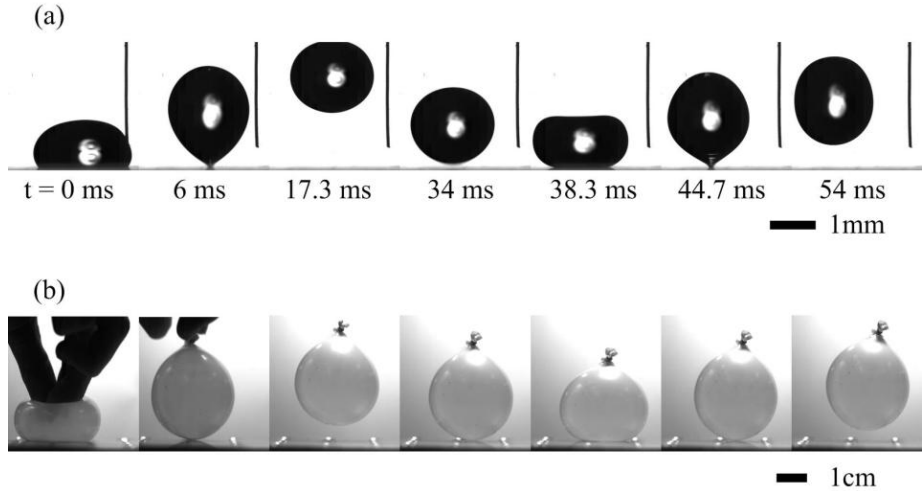

**Figure 2.** (a) Jumping and bouncing of a droplet sitting on superhydrophobic surface when driven by the resonant AC electrowetting: $100V_{RMS}$ and 39Hz. At t=0ms, the droplet is fully stretched, storing excessive surface energy. The surface energy is, then, converted to the kinetic energy for jumping, wherein the superhydrophobic surface provides very effective energy conversion. The potential energy of the airborne droplet at t=17.3ms shows that the energy conversion efficiency is fairly acceptable. (b) Jumping and bouncing of a water balloon show a same jumping mechanism as the droplet jumping, when it is pressed and released by hand.